\documentclass[12pt]{article}
\usepackage{epsf,amsmath}
\begin{document}

\title{Response to Comments by Myrvold and Appleby}

\author{Karl Hess$^1$ and Walter Philipp$^2$}

\date{$^1$ Beckman Institute, Department of Electrical Engineering
and Department of Physics,University of Illinois, Urbana, Il 61801
\\ $^{2}$ Beckman Institute, Department of Statistics and Department of
Mathematics, University of Illinois,Urbana, Il 61801 \\ }
\maketitle

\begin{abstract}

Myrvold and Appleby claim that our model for EPR experiments is
non-local and that previous proofs of the Bell theorem go through
even if our setting and time dependent instrument parameters are
included. We show that their claims are false.

\end{abstract}

We first review briefly the parameter space introduced by Bell
\cite{bellbook} and our extension of this parameter space. Then we
discuss the claims of Myrvold \cite{myr} and Appleby \cite{apy}.
We use the notation of our previous papers \cite{hp},
\cite{hpp1}-\cite{hpp3}.

Bell's parameter random variables are essentially given by the
functions ${A_{\bf a}}(\lambda) = \pm1, {B_{\bf b}}(\lambda) =
\pm1$ that indicate the spin value, with $\lambda$ being a source
parameter. It is generally assumed that the way EPR- experiments
are performed guarantees that $\lambda$ is independent of the
instrument settings ${\bf a}, {\bf b}$.

We add in our model \cite{hp}-\cite{hpp3} setting and time
dependent instrument parameter random variables $\lambda_{{\bf a},
t}^*$ operating at station $S_1$ and $\lambda_{{\bf b}, t}^{**}$
at station $S_2$.

These variables $\lambda_{{\bf a}, t}^*$, $\lambda_{{\bf b},
t}^{**}$ can be thought of as being generated by two computers
with equal internal computer clock time but otherwise entirely
independent. The ``values" that these variables assume could be
represented by any programs that evaluate the input of ${\bf a},
t$ etc. We do not claim knowledge of any mathematical properties
of these parameters as dictated by physics nor do we claim that
they must exist in nature. We can currently not simulate the EPR
experiment on such independent but time correlated computers and
never have claimed that we can. However, we postulate that any
proof of Bell-type inequalities that is relevant to locality
questions must pass the test to include setting and time dependent
instrument parameters. These instrument parameters do obey
Einstein locality and, therefore, must be covered by any EPR model
that is constructed like Bell's. We have shown, however, in
references  \cite{hp}-\cite{hpp3} that the standard proofs of
Bell-type inequalities \cite{bellbook} do not pass this crucial
test.

The reason for the difficulties of Bell-type proofs with such
parameters is that two time correlated computers can produce
virtually arbitrary setting and time dependent joint probability
distributions for Einstein local parameters $\lambda_{{\bf a},
t}^*$ and $\lambda_{{\bf b}, t}^{**}$ such as:
\begin{equation}
{\rho_s}(\lambda_{{\bf a}, t}^*, \lambda_{{\bf b}, t}^{**},
\lambda, m) \label{apy1}
\end{equation}
where $m$ corresponds to a concatenation of very short time
intervals that can be effected in various ways \cite{foot1}. The
subscript $s$ of $\rho$ indicates the setting dependence.

Bell and all his followers exclude such parameters since they
assume that their probability distribution $\rho(\lambda)$ does
not depend on any setting. The instrument parameters that they do
include are conditionally independent given $\lambda$. In other
words, if instrument parameters are included, then all Bell-type
proofs (see e.g. p 36 of \cite{bellbook}) assume a product
distribution:
\begin{equation}
{\rho_s}(\lambda_{{\bf a}, t}^*, \lambda_{{\bf b}, t}^{**}, m|
\lambda) = {\rho_1}(\lambda_{{\bf a}, t}^*, m
|\lambda)\cdot{\rho_2}(\lambda_{{\bf b}, t}^{**}, m|\lambda)
\label{apy2}
\end{equation}
Because we consider time correlations of the parameters, this
cannot be true. Thus Bell and followers exclude a large set of
joint probability densities. This was overlooked by Myrvold
\cite{myr} who falsely claimed that Bell-type proofs do go forward
with time and setting dependent instrument parameters. We have
already previously \cite{mg} written in more general terms about
other claims of Myrvold and turn now to the claims of Appleby
\cite{apy}.

Appleby claims that the functions $A_{\bf a}, B_{\bf b}$ of our
model violate parameter independence. Because we have $A_{\bf a} =
\pm1$ and $B_{\bf b} = \pm1$, parameter independence is equivalent
to:
\begin{equation}
E\{A_{\bf a}|\lambda\} = E\{B_{\bf b}|\lambda\} = 0 \label{apy3}
\end{equation}
where $E\{.|{\lambda}\}$ denotes the conditional expectation value
with respect to $\lambda$. Appleby claims that the corresponding
conditional expectation values of our work do not obey
Eq.(\ref{apy3}) and are setting dependent (see Eqs.(6)-(12) of
\cite{apy}). However, we have already shown in section 5.3 of
\cite{hp} that our model fulfills Eq.(\ref{apy3}) by an obvious
addition or extension namely the introduction of a Rademacher
function or any Lebesgue measurable function $r(t)$ that assumes
the values $\pm1$ on a set of measure $1/2$ each. We state in
\cite{hp}:

``To fulfill requirements of physics, it is necessary to be able
to obtain certain values ${-1} \leq {\alpha} \leq 1$ for
measurements on one side only and therefore one needs to be able
to have predetermined values for the following type of integrals
\begin{equation}
{\int}A {\rho} d u d v = \alpha \label{apy4}
\end{equation}
It is easily seen that this can be achieved without changing the
result for the pair correlation by use of functions $A, B$
generalized in the following way."

\begin{equation}
A_r := A_{\bf a}(u)r(z) \label{apy5}
\end{equation}
where $r(z)$ can be any Lebesgue measurable function that assumes
only values $\pm1$, and a similar equation for $B$. Then we wrote:
``The important special case $\alpha = 0$ is particularly easy to
achieve in a multitude of ways. For example one can choose a
function $r(t)$ (depending only on time t) that varies rapidly and
symmetrically between $\pm1$."

Appleby \cite{apy} did not follow our recipe to use $r(t)$. He
uses $r(\lambda)$ instead and obtains a useless result. Had he
used $r(t)$, it would have been obvious that one obtains
Eq.(\ref{apy3}) in our \cite{hp} model. If, on the other hand, one
deliberately makes the choice where $r = r(\lambda)$, then it is
equally obvious that Eq.(\ref{apy3}) may not hold as Appleby has
shown in great detail \cite{apy}.

For lack of space, we did not repeat our argumentation concerning
the Rademacher function in reference \cite{hpp2} as we were
confident that any reader would be able to apply Fubini's theorem
on double integration to the present context; in particular, when
the integral of Eq.(\ref{apy4}) is a product of two terms where
one of them $\int r(t) dt = 0$.

The same effect can be achieved in a simpler way. Instead of
introducing a new dimension, represented by the coordinate $t$, we
partition, in our original construction, each layer with label
$m$, where $m = 1,2,...,N$, into two parts say $(m, m')$ with $m,
m' = 1,2,...,N$. On the part with label $m$ everything remains the
same. On the part labelled $m'$ the density ${\rho}_{(m')} =
{\rho}_{(m)}$, remains the same, but $A_{\bf a}$ is replaced by
$-A_{\bf a}$ and $B_{\bf b}$ is replaced by $-B_{\bf b}$. Then for
each pair $(m, m')$
\begin{equation}
A_{\bf a}^{(m')} + A_{\bf b}^{(m)} = 0 \label{apy6}
\end{equation}
The expectation values of the $AB$ products remain unchanged.
Therefore, Appleby's claim of parameter dependence in our model is
invalid.

We are currently preparing a new and expanded version of our
construction that is hopefully clearer and somewhat simplified
while at the same time addressing all concerns so far published by
several authors.

\end{document}